\documentclass[12pt]{article}

\usepackage{epsfig}
\usepackage{graphicx}

\def\be{\begin{equation}}
\def\ee{\end{equation}}
\def\beq{\begin{eqnarray}}
\def\eeq{\end{eqnarray}}

\def\l{\lambda}

\def\r{\rho}

\def\vp{\varphi}

\def\N{{\cal N}}
\def\O{{\cal O}}

\def\({\left (}
\def\){\right )}
\def\[{\left [}
\def\[{\right ]}

\def\pr{{(\phi_{,r})}}

\begin{document}

\begin{titlepage}
\bigskip
\rightline{}
\rightline{hep-th/0406134}
\bigskip\bigskip\bigskip\bigskip
\centerline
{\Large \bf {Towards a Big Crunch Dual}}
\bigskip\bigskip
\bigskip\bigskip

\centerline{\large Thomas 
Hertog\footnote{Hertog@vulcan.physics.ucsb.edu} and 
Gary T. Horowitz\footnote{gary@physics.ucsb.edu}}
\bigskip\bigskip
\centerline{\em Department of Physics, UCSB, Santa Barbara, CA 93106}
\bigskip\bigskip

\begin{abstract}
We show there exist smooth asymptotically anti-de Sitter initial data 
which evolve to a big crunch singularity in a low energy supergravity limit of 
string theory. 
This opens up the possibility of using the dual conformal field
theory to obtain a fully quantum description of the cosmological singularity. 
A preliminary study of this dual theory suggests that the 
big crunch is an endpoint of evolution even in the full string theory. 
We also show that any theory with scalar solitons must have negative 
energy solutions. The results presented here clarify our earlier work on 
cosmic censorship violation in $N=8$ supergravity.

\end{abstract}

\end{titlepage}

\baselineskip=18pt

\setcounter{equation}{0}
\section{Introduction}

One of the main goals of quantum gravity 
is to provide a better understanding of the big bang or big crunch
singularities in cosmology. Perhaps the most fundamental question
is whether these singularities represent a true beginning or end of time
(perhaps described by a special quantum state) \cite{Hartle83}, 
or whether there
is some type of bounce, as envisioned by the pre-big bang \cite{Gasperini93}
or cyclic universe models \cite{Khoury02}.
Since our usual notions of space and time are likely to
break down near cosmological singularities, a particularly 
promising approach to study this issue might be
to find a dual description in terms of more fundamental variables.

In string theory we do not yet have a dual description of real cosmologies,
but we do
have the celebrated AdS/CFT correspondence \cite{Maldacena98} which provides 
a non-perturbative definition of string theory on asymptotically 
anti-de Sitter (AdS) spacetimes in terms of a conformal field theory (CFT). 
The dual CFT description has been used to study the singularity inside black 
holes \cite{Fidkowski04}, which is  analogous to a cosmological singularity. 
Although some progress in this direction has been made,  the fact
that the singularity is hidden behind an event horizon clearly complicates 
the problem. This is because the CFT evolution is dual to bulk evolution in 
Schwarzschild time so the CFT never directly `sees' the singularity.

It would be better to have examples of solutions in a low energy supergravity 
limit of string theory where smooth, asymptotically AdS initial data evolve to
a big crunch singularity. Then AdS/CFT should provide a precise framework in
which the quantum nature of cosmological singularities could be understood,
at least in asymptotically AdS spaces.
In this context, a big crunch singularity is simply any spacelike singularity 
which extends to infinity and reaches the boundary in finite time. 

In this paper we present examples of such solutions in the familiar $\N=8$, 
$D=4$ supergravity obtained by compactifying eleven dimensional supergravity 
on $S^7$. This theory has scalars with $m^2=-2$ in units of the AdS radius. 
It has recently been shown \cite{Hertog04} that there is a one parameter 
family of boundary conditions for this field which preserve the full set of 
asymptotic AdS symmetries. When this parameter vanishes, the dual CFT is the 
usual $2+1$ theory on a stack of M2-branes. Nonzero values of the parameter 
correspond to modifying this theory by a triple trace operator. We will show 
that for all nonzero values of the parameter, there are bulk solutions where
smooth, finite mass initial data evolve to a big crunch. We will also give
a preliminary argument that evolution in the dual CFT also ends in finite time.

As a second class of examples, we briefly discuss the case of $\N=8$, $D=5$ 
supergravity. This theory has scalars with $m^2=-4$ saturating the 
Breitenlohner-Freedman (BF) bound \cite{Breitenlohner82} in five dimensions. 
Once again, there is a one parameter family of AdS invariant boundary 
conditions \cite{Hertog04,Henneaux04}. 
We find that for almost all values of this parameter, there are solutions 
which evolve to a big crunch. $\N=8$ supergravity with these boundary 
conditions is similar to the theory discussed in 
\cite{Hertog03b} as a possible counterexample to cosmic censorship. 
Some valid criticism of that work was raised 
in \cite{Gutperle04} and \cite{Hubeny:2004cn},
but the results presented here can be viewed as confirmation
that with suitable boundary conditions, this theory
admits solutions where initial data evolve to singularities 
that are not hidden inside event horizons.

Having asymptotically AdS initial data evolving to a big crunch, is clearly a
sign of (nonlinear) instability. Other instabilities in the context of
AdS/CFT have been discussed previously \cite{Maldacena:1998uz,Hubeny:2004cn}.

We begin in the next section with 
a general result about stability. We show that any theory of gravity coupled 
to a scalar field that has static solitons must also have solutions with 
negative total energy. 
In section 3 we discuss our main example, including a description 
of the boundary conditions and a demonstration that $AdS_4$ is unstable. We 
also discuss what is known about the dual CFT and make some preliminary 
remarks about the dual field theory description of the big crunch.
Section 4 discusses the five dimensional example, and section 5 contains
some concluding remarks. In the appendix we generalize some of our calculations
to arbitrary dimensions and more general matter fields.

\setcounter{equation}{0}
\section{Scalar solitons imply negative energy}

Consider gravity coupled to a scalar field with potential $V(\phi)$ in
any spacetime dimension greater than three. 
In this section we prove the following result.

{\bf Claim:} If there exists a nonsingular, static, spherically symmetric,
finite
energy solution, then the positive energy theorem must be violated. 

In other words, if the theory has solitons, it must have negative energy
solutions. Note that we are not claiming that the soliton itself must
have negative energy, but only that negative energy solutions must
exist. Partial versions of this result applying to non-negative potentials
have already appeared in the literature.
In particular, the argument below is a slight generalization of the
scaling argument given in \cite{Heusler92}.

We begin with the result that a static solution must extremize the energy.
In the context of general relativity, this was shown in detail
by Wald and Sudarsky
\cite{Sudarsky:ty}.
Their argument is basically the following. The Hamiltonian $H$ is
a function of the spatial metric $g_{ij}$, scalar field configuration $\phi$,
and their conjugate momenta $\pi^{ij}$ and $p$. It takes 
the form of a volume integral of the constraints, plus a surface term.
For any solution, the volume term vanishes and the surface term is the total 
mass.
By definition, the variation of $H$ gives the time derivatives of the
fields. More precisely, for any perturbation $\delta g_{ij}, \delta \pi^{ij},
\delta \phi, \delta p$ one has
\be
\delta H = \int -\dot \pi^{ij} \delta g_{ij}+  \delta \dot  g_{ij} \delta
\pi^{ij} -\dot p \delta \phi + \dot \phi \delta p
\ee
For a static solution, the time derivatives are all zero (assuming
we choose the lapse and shift to correspond to the timelike Killing field).
So the static solution is an extremum of $H$. If we choose the perturbation
to satisfy the linearized constraints, then the volume contribution to 
$\delta H$ vanishes. Therefore static 
solutions must extremize the surface term,
i.e., the total mass. 

The above argument was initially given for asymptotically flat spacetimes,
but it also applies to asymptotically AdS spacetimes which we will 
explicitly consider below. So we assume $V(\phi)$ has a negative
extremum which $\phi$ approaches asymptotically. 
To continue, we will need an explicit formula
for the mass of  spherically symmetric (and time symmetric) initial data
when the scalar field
has a profile $\phi(r)$. We will work in $d$  spatial dimensions.
In this case, the constraint equations reduce to
(with $8\pi G=1$)
\be\label{constr}
\ ^{d}{\cal R} =  g^{ij}\phi_{,i}\phi_{,j} + 2 V(\phi) 
\ee
The spatial metric can be written as
\be \label{metric}
ds^2 = \left(1-{m(r)\over (d-1)r^{d-2}}+{r^2 \over \ell^2}
\right)^{-1} dr^2 + r^2 d\Omega_{d-1}, 
\ee
where $\ell$ is the radius of curvature of the asymptotic AdS space.
It is related to the asymptotic value of the potential, $\Lambda$, by
\be\label{llambda}
\ell^2= -\frac{d(d-1)}{2\Lambda}.
\ee
The constraint (\ref{constr}) then yields the following equation for $m(r)$ 
\be \label{mscalar}
m_{,r} +\frac{1}{d-1}m(r)r\pr^2 = r^{d-1} 
\left[2(V(\phi)-\Lambda)+\left(1+ {r^2\over \ell^2}\right)\pr^2 \right]
\ee
The general solution for arbitrary $\phi (r)$ is
\be\label{gensoln} 
m(r) = \int_{0}^{r} e^{-{1\over d-1}
\int_{\tilde r}^r  d\hat r \ \hat r\pr^2} 
\left[2(V(\phi)-\Lambda) +\left(1+ {\tilde r^2 \over \ell^2}\right)
\phi_{,\tilde r}^2 \right] \tilde r^{d-1} d\tilde r.  
\ee
The total mass is proportional to the asymptotic value of $m(r)$:
\be \label{totm}
M = {\pi^{{d \over 2}}\over \Gamma \left({d \over 2}\right)} \lim_{r\to\infty} m(r)
\ee

Now suppose $\phi_0(r)$ is a static soliton and consider the one parameter
family of configurations $\phi_\l(r) = \phi_0(\l r)$. Then from 
(\ref{gensoln}) and (\ref{totm}),
it is easy to see that the total mass takes the form
\be
M_\l = \l^{-d} M_1 + \l^{-(d-2)} M_2
\ee
where $M_2$ is independent of the potential and is manifestly positive, and
both $M_i$ are independent of $\l$. Since the soliton extremizes the energy
\be
0={d M_\l\over d\l}|_{\l=1} = -d M_1 -(d-2)M_2
\ee
Therefore, for all $d>2$ one has $M_1=-{d-2 \over d} M_2 <0$. 
The potential contribution to the mass is negative.
This means that the configuration $\phi_\l(r)$ with small $\l$ has negative
total mass. This argument clearly generalizes to several scalar fields and
general non-linear sigma model type kinetic terms: $G_{ab}(\phi) \nabla_\mu
\phi^a \nabla^\mu \phi^b$.

\setcounter{equation}{0}
\section{Main example}

Our main example starts with the low energy limit of string theory
with $AdS_4 \times S^7$ boundary conditions. The massless sector 
of the compactification of $D=11$ supergravity on $S^7$ is 
${\cal N}=8$ gauged supergravity in four dimensions \cite{deWit82}.
The bosonic part of this theory
involves the graviton, 28 gauge bosons in the adjoint of $SO(8)$,
and 70 real scalars, and admits $AdS_4$ as a vacuum solution.
It is possible to consistently truncate this theory to  include only 
gravity and a single scalar with action \cite{Duff99}
\be\label{4-action}
S=\int d^4x\sqrt{-g}\left[\frac{1}{2}R
-\frac{1}{2}(\nabla\phi)^2 +2+\cosh(\sqrt{2}\phi) \right]
\ee
(We have chosen the gauge coupling so that the AdS radius is equal to one.)
Notice that the potential is unbounded from below, and the scalar has mass
\be
m^2 = -2 \ .
\ee
Even though the field is tachyonic, the mass is above the 
Breitenlohner-Freedman bound $m_{BF}^2 = -9/4$ \cite{Breitenlohner82}, 
and with the usual boundary conditions $AdS_4$ is stable.

\subsection{AdS Invariant Boundary Conditions}

We will mainly work in global coordinates in which the $AdS_4$ metric
takes the form
\be \label{adsmetric}
ds^2_0 = \bar g_{\mu \nu} dx^{\mu} dx^{\nu}=
-(1+r^2 )dt^2 + {dr^2\over 1+r^2} + r^2 d\Omega_{2}
\ee
For $m^2 =-2$, solutions to the linearized wave equation
$\nabla^2\phi -m^2\phi=0$ with harmonic time
dependence $e^{-i\omega t}$ all fall off asymptotically like 
\be\label{genfall}
\phi = {\alpha \over r}  + {\beta \over r^2}
\ee
The standard boundary conditions correspond to 
taking $\alpha=0$ in (\ref{genfall}), so the modes fall off as fast as 
possible.
For those boundary conditions, the  total energy is finite, using
the standard definition of energy \cite{Abbott82}. 
Recall that the energy, and more generally, conserved charges associated
with asymptotic symmetries $\xi^\mu$ can be defined as follows 
\cite{Henneaux85}.  One starts with the
Hamiltonian $H[\xi] = \int d^3x \xi^\mu H_\mu$ where $H_\mu$ are the
usual constraints, adds surface terms so that $H$ has well defined
functional derivatives, and then subtracts the analogous expression for
the $AdS_4$ background. This gives
\be \label{gravcharge}
Q_G[\xi]= \frac{1}{2}\oint dS_i
\bar G^{ijkl}(\xi^\perp \bar{D}_j h_{kl}-h_{kl}\bar{D}_j\xi^\perp)
+2\oint dS_i\frac{\xi^j {\pi^i}_j}{\sqrt{\bar g}}
\ee
where $G^{ijkl}={1 \over 2} g^{1/2} (g^{ik}g^{jl}+g^{il}g^{jk}-2g^{ij}g^{kl})$,
$h_{ij}=g_{ij}-\bar{g}_{ij}$ is the deviation from the spatial metric 
$\bar{g}_{ij}$ of pure AdS, $\bar{D}_i$ denotes covariant differentiation 
with respect to $\bar{g}_{ij}$ and $\xi^\perp = \xi^{\mu} n_{\mu}$ with
$n_{\mu}$ the unit normal to the surface. 

However since $m^2 =-2$ lies in the range $ m^2_{BF} +1 >m^2 >m^2_{BF}$ 
there exists an additional one-parameter family of AdS invariant boundary
conditions on the scalar field and the metric components \cite{Hertog04}. 
More precisely, the asymptotic AdS symmetries are preserved in solutions that
belong to the following class,
\be 
\label{4-scalar}
\phi(r,t,x^{a})=\frac{\alpha (t,x^a)}{r}+
\frac{f \alpha^2 (t,x^a)}{r^2}
\ee
\beq 
\label{4-grr}
g_{rr}=\frac{1}{r^2}-\frac{(1+\alpha^2/2)}{r^4}+
O(1/r^5) & \quad g_{tt}=-r^2 -1+O(1/r) \nonumber\\
g_{tr}=O(1/r^2) \qquad \qquad \qquad & \ \  \ \ \ g_{ab}= \bar g_{ab} +O(1/r) 
\nonumber\\
g_{ra} = O(1/r^2) \qquad \qquad \qquad & g_{ta}=O(1/r) \ \ 
\eeq
where $x^a=\theta,\phi$ and $f$ is an arbitrary constant 
that labels the different boundary conditions. Notice that the
boundary conditions on some of the metric components are also relaxed compared
to the standard set. When $f=0$ we recover the $\beta=0$ boundary conditions 
which have been considered before in the context of AdS/CFT \cite{Klebanov99}. 
Remarkably, 
however, the asymptotic anti-de Sitter symmetries are preserved for all 
values of $f$. In particular, it is easy to see that
rescaling $r$ leaves $f$ unchanged. Since $\alpha$ depends on the
particular solution and can vanish, each of these boundary conditions admits
$AdS_4$ as a solution.

For these more general boundary conditions, the usual energy 
(\ref{gravcharge}) diverges. However,
one can define finite conserved charges by repeating the above 
procedure. The net result is that there is an additional 
surface term involving the scalar field\footnote{The form of
the surface term given here is different from that in \cite{Hertog04}, but it
is equivalent. Both the divergent and finite terms agree. On the other hand,
the finite contribution differs from the energy defined by  
holographic renormalization, as discussed in \cite{deHaro:2000xn,Bianchi02}, 
since they require that the energy vanishes for a nontrivial
domain wall background.} \cite{Hertog04}
\be
\label{charge4d}
Q[\xi]=Q_G[\xi]+{1 \over 6} \oint \xi^\perp 
\left[ (\nabla \phi)^2 -m^2\phi^2 \right]    
\ee
For all finite $f$ (including $f=0$!) the scalar and gravitational charge 
separately diverge. The divergences, however, exactly cancel out yielding
finite total charges $Q[\xi]$. 
By contrast, the scalar charges $Q_\phi$ 
vanish for the standard boundary conditions (\ref{genfall}) with $\alpha=0$.

For spherical solutions it is easy to compute the total mass $M=Q[\partial_{t}]$,
which yields
\be \label{mass4dhair}
M=4\pi \left( M_0+\frac{4}{3}f\alpha^3 \right).
\ee
where $M_0$ is the coefficient of the $O(1/r^5)$ correction to $g_{rr}$.

\subsection{Solitons}

We now look for static, spherically symmetric 
asymptotically AdS soliton solutions. Writing the metric as
\be
ds^2=-h(r)e^{-2\delta(r)}dt^2+h^{-1}(r)dr^2+r^2d\Omega_2
\ee
the field equations read
\be\label{hairy14d}
h\phi_{,rr}+\left(\frac{2h}{r}+\frac{r}{2}\phi_{,r}^2h+h_{,r}
\right)\phi_{,r}   =  V_{,\phi}
\ee
\be\label{hairy24d}
1-h-rh_{,r}-\frac{r^2}{2}\phi_{,r}^2h =  r^2V(\phi)
\ee
\be
\delta_{,r} = -{ r \phi_{,r}^2 \over 2}
\ee
Regularity at the origin requires $h=1$, $h_{,r}=0$ and $\phi_{,r}=0$ at $r=0$.

\begin{figure}[htb]
\begin{picture}(0,0)
\put(34,258){$\phi$}
\put(440,21){$r$}
\end{picture}
\mbox{\epsfxsize=15cm \epsffile{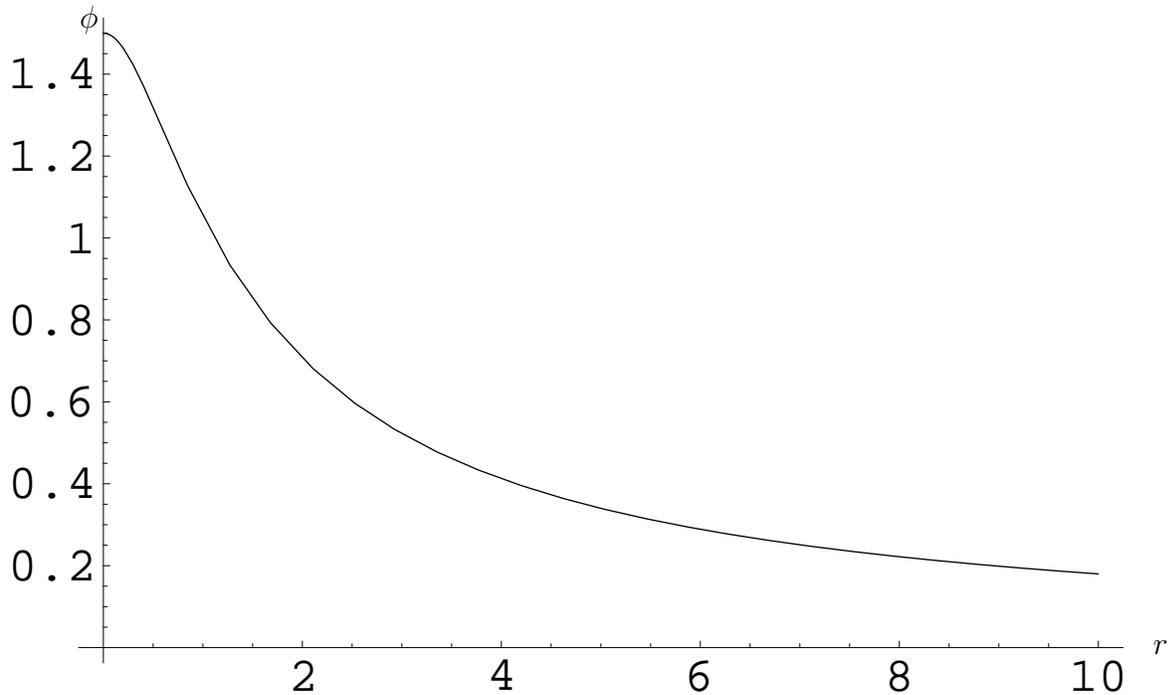}}
\caption{Soliton solution $\phi(r)$ 
with boundary conditions specified by $f=-1/4$.}
\label{1}
\end{figure}

We have numerically solved these equations.
For every nonzero $\phi_0$ at the origin, the solution
to (\ref{hairy14d}) is asymptotically of the form (\ref{4-scalar}) 
for some value of $f$. The staticity and spherical symmetry of the soliton 
mean $\alpha (t,x^a)$ is simply a constant.
The scalar field value $\phi_0$ at the origin uniquely 
determines $f$ and vice versa; there is at most one static spherical soliton 
solution in each theory. We find a regular soliton solution for all finite 
$f \neq 0$. When $|f| \rightarrow 0$ we find 
$ |\phi_0|  \rightarrow \infty$ and
for $|f| \rightarrow  \infty$ one has $ |\phi_0|  \rightarrow 0$ so the
nontrivial soliton solution ceases to exist in this limit.
As an example, in Figure 1 we show the soliton solution for
$f=-.25$ boundary conditions, which has $\phi_0 \approx 1.5$. 

The existence of soliton solutions for a large class of AdS invariant 
boundary conditions implies there are negative mass solutions
in those theories. This is because the soliton solutions extremize the 
energy (\ref{mass4dhair}). Although the total mass $M$ now acquires
a contribution from the scalar field, the argument given in Section 2 
still applies: the total mass consists of a finite term $M_1$ that scales
as the volume (which includes the scalar contribution) and a finite,
positive term $M_2$ that scales linearly in $r$. Hence $M_1$ must be
negative for the soliton, which means rescaled configurations
$\phi_\l(r) = \phi_0(\l r)$ with sufficiently small $\l$ must have 
negative mass. 

Such rescaled configurations are initial data for time-dependent solutions.
By adjusting $\l$, one can arrange to have an arbitrarily large
central region where $\phi$ is essentially constant and 
away from the maximum of the potential. It follows that the field must
evolve to a spacelike  singularity \cite{Hertog03b}. 
Moreover, the singularity
that develops cannot be hidden behind an event horizon, because all 
spherically symmetric black holes have positive mass \cite{Hertog04}.
Instead, one expects 
it to continue to spread, cutting off all space.\footnote{If $V$ were bounded
from below, it has been shown that the singularity cannot end or become
timelike \cite{Dafermos:2004ws}. The same is likely to be true here.} 
The existence of solitons, therefore, indicates the existence of 
finite mass configurations that produce a big crunch.
A particular example of such a configuration for which this can be shown
explicitly is given next.

\subsection{Instantons}

First we construct an $O(4)$-invariant Euclidean instanton solution of the form
\be \label{inst}
ds^2 = {d\r^2\over b^2(\r)} +\r^2 d\Omega_3
\ee
and $\phi=\phi(\r)$. The field equations determine $b$ in terms of $\phi$
\be \label{inst1}
b^2(\r) = { 2V \r^2 -6\over \r^2 \phi'^2 -6}
\ee
and the scalar field $\phi$ itself obeys
\be\label{inst2}
b^2 \phi'' + \left( {3 b^2 \over \r} +bb' \right) \phi' -V_{,\phi} =0
\ee
where prime denotes $\partial_{\rho}$.
Regularity again requires $\phi'(0) =0$.

\begin{figure}[htb]
\begin{picture}(0,0)
\put(37,263){$\phi$}
\put(440,21){$r$}
\end{picture}
\mbox{\epsfxsize=15cm \epsffile{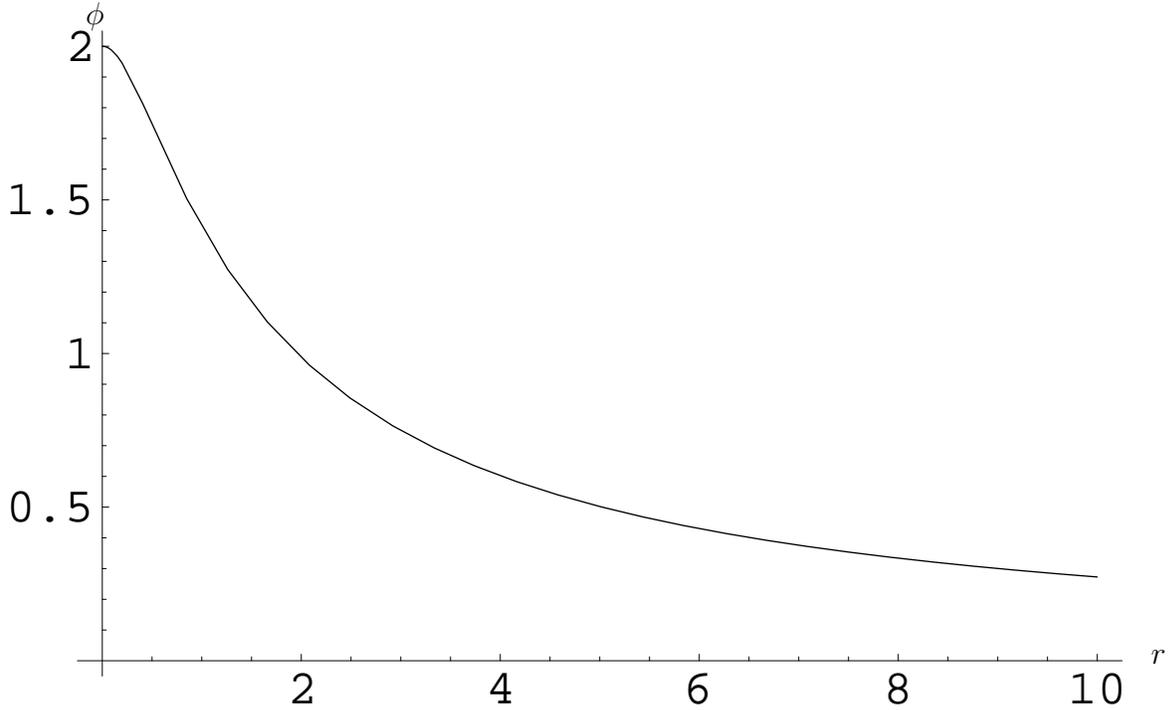}}
\caption{Instanton solution $\phi(\r)$ 
with boundary condition $f=-1/4$.}
\label{2}
\end{figure}

 From (\ref{inst2}) it follows that asymptotically $\phi(\r)$ has the same
behavior as the Lorentzian scalar field solutions considered above,
\be \label{as-scalar}
\phi = {\alpha\over \r} + {f \alpha^2\over \r^2}.
\ee
We find that all boundary conditions that admit a spherical soliton solution
also admit an $O(4)$-invariant instanton solution. As for the solitons, 
 $f$ is determined by the field $\phi(0)$ at the origin. 
In Figure 2, the profile $\phi(\r)$ is shown of the instanton 
with $f=-.25$ boundary conditions.

The solution we want is obtained by analytically continuing the instanton.
We will discuss this in the next subsection, but first we comment on
the interpretation of this instanton.
The existence of negative energy solutions indicates that $AdS_4$ is
unstable with these boundary conditions.
The quantum decay rate is determined in a
semiclassical approximation by the Euclidean action of instanton. 
 The action is given by
\be
I = \int [-{1\over 2} R + {1\over 2}(\nabla \phi)^2 + V(\phi)]
  - \oint K +{1 \over 6} \oint \left[ (\nabla \phi )^2 -m^2\phi^2 \right]    
\ee
where the first surface term is the usual Gibbons-Hawking term, and the
second is the surface term required so that the Hamiltonian constructed
from this action (after subtracting the background) agrees with 
(\ref{charge4d}).
For the case $f=0$ this term is equivalent to the surface term 
$-{1 \over 2} \oint \phi \nabla_i \phi dS^i $ introduced by 
Klebanov and Witten to render the action of the $\alpha/r$ modes finite 
\cite{Klebanov99}.

The relevant quantity for computing the rate of vacuum decay is the
difference between the instanton action and the action for pure AdS:
$\Delta I = I - I_{AdS}$. Subtracting $I_{AdS}$ removes the leading
divergences in $I$, but since $\phi$  goes to zero so slowly,
there are two subleading divergences. If the coefficients of these terms
were not exactly zero, $\Delta I$ would be infinite and there would
be no probability for the vacuum to decay. In appendix A 
we show that both coefficients miraculously 
vanish. This involves nontrivial cancellations among the
volume term and both surface terms in the action. 
Furthermore, the difference $\Delta I$ becomes small for large $\vert f \vert$
and goes to zero when $\vert f \vert \rightarrow \infty$.

\subsection{Big Crunch Instability}

Consider the slice through the instanton obtained by restricting to the
equator of the $S^3$. The fields on this surface define time symmetric
initial 
data for a Lorentzian solution. The Euclidean radial distance $\r$ simply
becomes the radial distance $r$ in the Lorentzian solution.
The total mass (\ref{mass4dhair})
of this initial data can be computed from the instanton geometry.
Substituting (\ref{as-scalar}) into (\ref{inst1}) yields asymptotically
\be\label{as-b}
b^2(\r) = \r^2  +1 +{\alpha^2 \over 2} + {4f\alpha^3\over 3 \r}
\ee
This is of the form (\ref{4-grr}) required to have 
finite conserved charges. In fact, we see that $M_0= -4f\alpha^3/3$ and hence
(\ref{mass4dhair}) implies that the total mass is zero!
 This is
consistent with its interpretation as the solution $AdS_4$ decays into.

The evolution of this initial data is simply obtained by analytic
continuation of the instanton. This is discussed in detail in
\cite{Coleman:1980aw}, but the basic idea is the following.
The origin of the Euclidean instanton
becomes the lightcone of the Lorentzian solution. Outside the
lightcone, the solution is given by (\ref{inst}) with $d\Omega_3$ replaced by
three dimensional de Sitter space. The scalar field $\phi$ remains bounded in
this region. Inside the lightcone, the $SO(3,1)$ symmetry ensures that
the solution evolves like an open FRW universe,
\be \label{ametric}
ds^2 = -dt^2+ a^2(t) d\sigma_3
\ee
where $d\sigma_3$ is the metric on the three dimensional unit hyperboloid.
The field equations are
\be\label{fieldeq}
{\ddot a\over a}= {1\over 3}[V(\phi) - \dot\phi^2]
\ee
\be\label{phieq}
\ddot{\phi} +{3 \dot a\over a} \dot\phi +V_{,\phi} =0 
\ee
and the constraint equation is
\be\label{constraint}
\dot a^2 - {a^2\over 3}\left [{1\over 2} \dot \phi^2 + V(\phi)\right ] = 1 \ ,
\ee
where $\dot a =\partial_{t} a$.
On the light cone, $\phi=\phi(0)$ and $\dot \phi=0$ (since $\phi_{,\r}=0$
at the origin in the instanton). 
Under evolution $\phi$ rolls down the negative potential, so the right
hand side of (\ref{fieldeq}) decreases. This ensures that $a(t)$ vanishes
in finite time
producing a big crunch singularity. For the purpose of understanding
cosmological singularities in string theory, one can forget the
origin of this solution as the analytic continuation of an instanton.
We have simply found an explicit example of asymptotically AdS initial
data which evolves to a big crunch.

The discussion above is closely analogous to the situation in theories
with a true vacuum as well as a false vacuum, in which one can describe
the decay of the false vacuum by nucleating a bubble of true vacuum.
One typically finds that small bubbles of true vacuum collapse, large
bubbles expand, and a critical size bubble is static. Our soliton
is like a static critical bubble. The solution obtained from the instanton
is like a larger bubble which expands. This can be seen by comparing 
Fig. 1 and Fig. 2. The
field profile obtained from the instanton is indeed
broader than the static soliton.

\subsection{Dual CFT description}

Having shown that the bulk theory admits solutions which evolve to
a big crunch, we now turn to the dual CFT description of this theory.
The dual to string theory on $AdS_4\times S^7$ can be obtained by
starting with the field theory on a stack of $N$ D2-branes. This is a
$SU(N)$ gauge theory with seven adjoint scalars $\vp^i$. One then takes
the infrared (strongly coupled) limit to obtain the CFT. In the process,
one obtains an $SO(8)$ symmetry. In the abelian case, $N=1$, this can be
understood by dualizing the three dimensional gauge field to obtain another
scalar. But in general, it is not well understood. 

This theory has dimension one operators $\O_T=Tr T_{ij} \vp^i\vp^j$ where
$T_{ij}$ is symmetric and traceless \cite{Aharony:1998rm}.\footnote{Since there are only seven
$\vp$'s and the theory has $SO(8)$ symmetry, there are other operators
involving the gauge field which complete the $SO(8)$ representation.}
One of these, $\O$,
is dual to the bulk field we
have been considering with the boundary conditions that
$\phi = \alpha/r +O(r^{-3})$ for physical states. The field theory
dual to the ``standard" quantization, where physical states are described by
modes with $\phi =\beta/r^2$ asymptotically, can be obtained by adding
the double trace term ${f\over 2}\int \O^2$ to the action
\cite{Witten02,Gubser:2002vv}.
This is a relevant perturbation and the infrared limit is another CFT
in which $\O$ has dimension two. 

As described in \cite{Hertog04}, the AdS invariant boundary conditions 
correspond instead to adding a triple trace term to the action
\be\label{Ocubed}
S = S_0 + {f\over 3} \int \O^3
\ee
This follows from Witten's treatment of multi-trace operators in AdS/CFT
\cite{Witten02}.
The extra term has dimension three, and hence is marginal and preserves
conformal invariance, at least to leading order. One might wonder if
this symmetry is exact, or whether
the operator $\O^3$ has an anomalous dimension. The anomalous dimension
can receive contributions proportional to $1/N$ or $f$. Since the large $N$
limit corresponds to supergravity in the bulk with AdS invariant boundary
conditions, and for every $f$ there is a bulk solution corresponding to 
pure AdS, it seems likely that the theory remains conformally invariant
for finite $f$ (at least for large $N$).\footnote{This is different from the
example considered in \cite{Gubser03} which also had a bulk AdS solution, but 
imposed boundary conditions that were not AdS invariant.}

Since the Lorentzian solution obtained from the instanton takes the form
(\ref{inst}) with $d\Omega_3$ replaced by
three dimensional de Sitter space, $dS_3$,
one might think that the natural dual would
correspond to the CFT on  $dS_3$. This
field theory certainly allows evolution for infinite time and is
nonsingular. But this only corresponds to evolution for finite global time.
We want to conformally rescale
$dS_3$ to (part of) the cylinder $R\times S^2$. This is equivalent to
a coordinate transformation in the bulk. The relation between the usual
static coordinates (\ref{adsmetric}) for $AdS_4$ and the $SO(3,1)$ invariant 
coordinates
\be
ds^2 = {d\r^2\over 1+\r^2} + \r^2 (-d \tau^2 + \cosh^2\tau d\Omega_2)
\ee
is 
\be
\r^2 = r^2 \cos^2 t -\sin^2 t
\ee
Since our bulk solution asymptotically has
\be
\phi(\r) = {\alpha\over \r} + {f\alpha^2\over \r^2}+ O(\r^{-3})
\ee
This becomes
\be
\phi(r) = {\tilde \alpha\over r} + {f\tilde\alpha^2\over r^2} +O(r^{-3})
\ee
where $\tilde\alpha= \alpha/\cos t$. Notice that $f$ is unchanged.
Hence the evolution of the initial data that we described in section 3.4
preserves the AdS invariant boundary conditions (\ref{4-scalar})-(\ref{4-grr}).
This also provides some support for the fact that (\ref{Ocubed}) is
conformally invariant for finite $f$. The fact that $\tilde\alpha$ blows
up as $t\rightarrow \pi/2$ is consistent with the fact that this is the
time that the big crunch singularity hits the boundary. Since the 
coefficient of $1/r$ is usually interpreted as the expectation value of $\O$
in the CFT, this also indicates that the CFT on the cylinder does NOT
have well defined evolution for all time. 

A qualitative explanation for this is the following.
The term we have added to the action is not positive definite. Since the
energy associated with the asymptotic time translation in the bulk can
be negative, the dual field should also admit negative energy states. This 
strongly suggests that the usual vacuum is  unstable. It might
decay via the (nongravitational) decay of the false vacuum. 
Perhaps a useful analogy is
a scalar field theory with potential $V= m^2 \vp^2 - f\vp^6$. The quadratic
term is analogous to the  coupling of $\vp$ to the curvature of $S^2$, which
is needed for conformal invariance. The second term is analogous to the
second term in (\ref{Ocubed}). Qualitatively
this theory has the same behavior as the bulk. There
are instantons which describe the semiclassical decay of the usual
vacuum $\vp=0$. For small $f$, the potential barrier is large, and the
instanton action is large. So tunneling is suppressed. For large $f$,
the barrier is small and tunneling is not suppressed. After the
tunneling, the field rolls down the potential and becomes infinite in 
finite time. So in the semiclassical description of this analogous field
theory, evolution ends in finite time. If the same is true in
the full description of the theory (\ref{Ocubed}) one could conclude that
there is no bounce through the big crunch singularity in the bulk. 

We close this section with a few comments about generalizations.
We have focused in this section on big crunch singularities, 
but the solutions we have presented have a big bang
singularity in the past as well, which should have a dual CFT description too. 
In addition,  it is likely there exist solutions with only one singularity, 
in the future or the past. We have also restricted ourselves to gravity 
coupled to a single scalar, but this was just for simplicity. We expect that 
with suitable boundary conditions, there are more general bulk solutions to 
low energy string theory that form a big crunch which include several scalars,
and possibly other matter fields. 

\setcounter{equation}{0}
\section{A Big Crunch in $D=5$ Supergravity}

${\cal N}=8$ gauged supergravity in five dimensions
\cite{Gunaydin85,Pernici85} is thought to be a consistent 
truncation of ten dimensional type IIB supergravity on $S^5$. The spectrum of 
this compactification involves 42 scalars parameterizing the coset 
$E_{6(6)}/USp(8)$. We will focus on a restriction of this theory
 to only gravity and one scalar \cite{Freedman:1999gk}. The action is
\be \label{lagr}
S = \int d^5x \sqrt{-g}\left[ \frac{1}{2} R -\frac{1}{2}(\nabla \phi )^2 
+\(2e^{2\phi/\sqrt{3}}_{\ }+4e^{-\phi/\sqrt{3}}_{\ } \)\right ]
\ee
(We have again adjusted the gauge coupling so that the AdS radius is one.)
As before, the potential is unbounded from below. Since the analysis
in this section is similar to that in the previous section, our discussion
will be brief.

The scalar $\phi$ has mass
\be
m^2= -4
\ee
Since this saturates the BF bound in five dimensions, $\phi$
asymptotically behaves as
\be \label{as-sc2}
\phi (t,r,x^a)= {\alpha (t,x^a) \ln r\over r^2} 
                + {\beta (t,x^a) \over r^2}
\ee
In \cite{Hertog04,Henneaux04} it was shown that the asymptotic AdS symmetry 
group is preserved for solutions with the following asymptotic behavior,
\be
\beta = \alpha \left(f-\frac{1}{2}\ln \alpha \right)
\ee
together with appropriate boundary conditions on the metric components.
Here $f$ is again an arbitrary constant.
For solutions within this class there exists a well-defined notion of
mass, namely the conserved charge $M=Q[\xi]$ associated with the
asymptotic Killing vector $\xi = \partial_t$. As before, this is finite but
acquires a contribution from the scalar field. 
For spherical solutions it is given by (see (\ref{genmass5d}) in the appendix)
\be \label{mass5d}
M=2\pi^2 \left[{3 \over 2}
M_0+\frac{1}{4}\alpha^2(\ln \alpha)^2
+\alpha^2\left(\frac{1}{4}-f\right)\ln \alpha
+\alpha^2\left(f^2-\frac{1}{2}f+\frac{1}{8}\right)\right].  
\ee
where $M_0$ is now the coefficient of the
$O(1/r^6)$ correction to the $g_{rr}$ component of 
the metric.

\begin{figure}[htb]
\begin{picture}(0,0)
\put(34,258){$\phi$}
\put(440,21){$r$}
\end{picture}
\mbox{\epsfxsize=15cm \epsffile{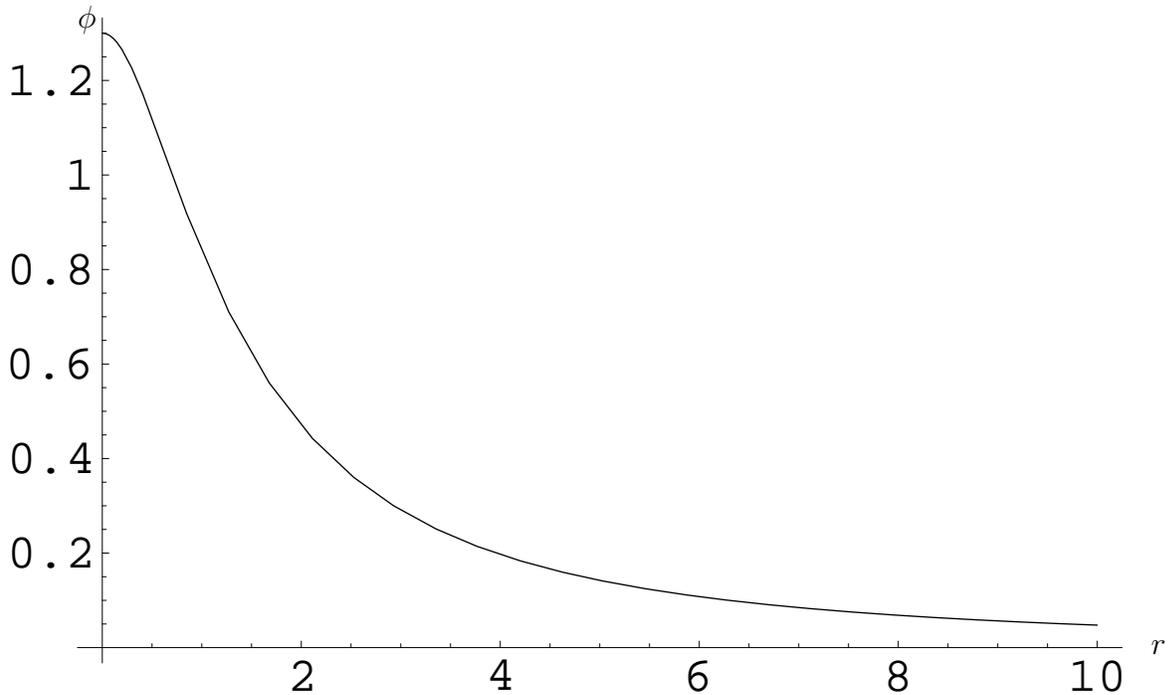}}
\caption{Instanton solution $\phi(\r)$ with boundary condition $f=1$.}
\label{3}
\end{figure}

To see that there are solutions which evolve to a big crunch we again start
with the instanton.
For finite $f$ boundary conditions (which allow a logarithmic scalar mode)
there exist regular $O(5)$-invariant Euclidean instanton solutions,
\be \label{inst5d}
ds^2 = {d\r^2\over b^2(\r)} +\r^2 d\Omega_4
\ee
with $\phi=\phi(\r)$ a solution of
\be\label{inst5d2}
b^2 \phi'' + \left( {4 b^2 \over \r} +bb' \right) \phi' -V_{,\phi} =0
\ee
An example of an instanton solution with $f=1$ boundary conditions is shown
in Figure 3. All finite $f$ boundary conditions admit precisely one instanton 
solution. The analytic continuation of the instanton again describes
asymptotically AdS initial data evolving to a big crunch.
In the appendix we show that the total mass (\ref{mass5d}) of 
this initial data  vanishes. On the other hand, 
the mass of static, spherical black holes that are solutions with the same 
boundary conditions is positive \cite{Hertog04}. This shows that
 the singularity that 
develops from the initial data cannot
be hidden behind an event horizon. Indeed, the explicit solution shows that,
with generalized AdS invariant boundary conditions, 
the singularity reaches infinity in finite time.

If one requires $\alpha=0$ in (\ref{as-sc2}) the logarithmic mode is switched 
off. For this choice of boundary conditions we do not find an instanton.
There is, however, still a finite scalar contribution to the conserved 
charges\footnote{This is also evident in the spinorial proof \cite{Gibbons83} 
of the positive energy theorem for these boundary conditions, where the 
positive surface term (which equals $Q[\xi]$) contains an extra finite scalar 
contribution \cite{Hertog03b}.}. This means the standard gravitational 
mass that appears in the metric need not be positive, even though the 
scalar modes fall off as fast as possible.
Examples of initial data with negative 
gravitational mass were given in \cite{Hertog03b}. At first sight this 
suggests $\alpha=0$
boundary conditions may also admit solutions in which smooth, finite mass 
initial data evolve to a big crunch. However, the gravitational mass alone is 
generally not conserved during evolution \cite{Hertog03b} (this has also been 
observed in the numerical work of \cite{Gutperle04}). Instead, the quantity 
that is conserved during evolution is the total charge $Q[\partial_t]$.
This is always positive \cite{Gibbons83} and we have not found any example
where a big crunch is produced from smooth asymptotically AdS 
initial data with these boundary conditions.

We have previously argued \cite{Hertog03b} that cosmic censorship is 
generically violated in ${\cal N}=8$, $D=5$ supergravity when one allows the 
logarithmic mode for one or more scalars saturating the BF bound. We
claimed that one could produce singularities that were not contained
inside a black hole. Our argument consisted of trying to find 
 initial data that evolve
to a singularity in a central region, but do not have enough mass to form a 
black hole large enough to enclose the singular region\footnote{In 
\cite{Hertog03b} we imposed a large radius cutoff at $R_1$ and required 
$\phi(R_1)$ to be constant in time to regularize the gravitational mass. 
This enabled us to compare the mass of the initial data with the mass required
to form a (hairy) black hole obeying the same boundary condition at $R_1$. 
Working with a finite radius cutoff is of course somewhat artificial because 
this breaks asymptotic AdS invariance. The cutoff is not needed, however, 
for the class of solutions considered here, because the boundary conditions
are AdS invariant despite the logarithmic branch. The comparison of the mass,
therefore, can be done using the conserved charge (\ref{mass5d}). 
We note, however, that the 
evolution with fixed $f$ is likely to differ from the evolution with
fixed $\phi(R_1)$.}. Instead, we argued, 
the singularity must have a naked endpoint or extend all 
the way to infinity, like a big crunch. In both cases one might argue
that cosmic censorship is violated
since one cannot evolve for all time in the asymptotic 
region. Our conclusions have been questioned in \cite{Gutperle04} and 
\cite{Hubeny:2004cn}, and indeed the examples we gave in \cite{Hertog03b} do
have enough mass to form a black hole. Nevertheless, 
the results of this section show that our basic conclusion was essentially
correct. The analytic continuation of the instanton shows that,
for ${\cal N}=8$, $D=5$ supergravity with general AdS invariant boundary
conditions, there
are indeed smooth initial data which evolve to big crunch singularities.
This is perhaps less surprising in light of the fact that the positive 
energy theorem is violated with these boundary conditions.

From the current standpoint of understanding big crunch singularities,
this example is less satisfactory than the one discussed in the previous
section. This is because it
is not clear what the CFT deformations are that correspond to
general AdS invariant boundary conditions in this case. 
There is a dimension two operator which is dual to
our scalar field with boundary conditions (\ref{as-sc2}) with $\alpha =0$. 
A naive application of Witten's prescription \cite{Witten02}
involves taking the logarithm of this operator.

\setcounter{equation}{0}
\section{Conclusion}

Motivated by a desire to find a dual quantum description of a cosmological 
singularity, we have constructed low energy string theory solutions that 
describe the evolution of asymptotically AdS initial data to a big crunch. 
Our main example involves a negative $m^2$ scalar field in $\N=8$, $D=4$
supergravity. To induce the instability, we have modified the boundary 
conditions on the scalar at infinity\footnote{This is a nonlinear
instability since, at the linearized level, the boundary conditions 
(\ref{4-scalar}) reduce
to the $\beta =0$ case discussed in \cite{Breitenlohner82}.}.
Although the modified boundary 
conditions preserve the full set of asymptotic AdS symmetries and allow a 
finite conserved energy to be defined, this energy can be negative. We have 
shown this by explicitly constructing negative energy initial data 
(the rescaled solitons of section 3.2). We have also constructed finite 
action instantons in this theory which describe the semiclassical decay of 
$AdS_4$. Most importantly, the analytic continuation of the instanton 
describes smooth (zero energy) initial data evolving to a big crunch.

We have also given an example of a similar solution in $\N=8$ supergravity 
in five dimensions, involving a scalar field saturating the BF bound. 
More generally, we show in the appendix
that a big crunch can be produced from smooth 
initial data whenever one has a scalar field with $m^2_{BF}\le m^2<m^2_{BF}+1$
which decouples from the rest of the matter. 

Usually one discards unstable theories, saying they are not of physical
interest. Here, we are using the fact that there should be a dual CFT
description of these bulk theories even if they are unstable.
The field theory should provide a complete quantum 
description of the big crunch singularity. 
Clearly the next step is to try to better understand these dual theories.
If states in the CFT have
a well defined evolution for all time, and a semiclassical bulk metric
can be reconstructed at late time, then there must be a bounce through
the singularity. However, if the CFT evolution ends after finite time,
or a semiclassical metric cannot be constructed, then the bulk evolution
would end at the big crunch. 

In $\N=8$, $D=4$ supergravity we find solutions  which evolve to a
big crunch
for a one parameter family of AdS invariant boundary conditions. These
solutions do not exist when the
parameter vanishes, in which case the dual CFT is the 
usual $2+1$ theory on a stack of M2-branes. Nonzero values of the parameter 
correspond to modifying this field theory by a triple trace operator.
Since the energy associated with asymptotic time translations in the bulk 
can be negative, AdS/CFT predicts this dual field theory should also admit 
negative energy states. This is plausible, since the extra term we have
added to the CFT action is not positive definite. We have
given preliminary evidence that
for some states, the CFT does not have well defined
evolution for all time (at least in the large $N$ limit). 
The dual CFT corresponding to $\N=8$, $D=5$ supergravity with generalized
boundary conditions is understood even less well, but investigations
in a related context \cite{Hubeny:2004cn} indicate it 
exhibits similar instabilities. 

Furthermore, we have evidence that bulk solutions  cannot evolve to a
big crunch in supergravity theories that are dual to stable CFT's.
In \cite{Hertog03,Hertog04b} we  considered gravity coupled
to scalar fields with potential satisfying the positive energy
theorem (in AdS)\footnote{These potentials 
can always be derived from a superpotential
$W(\phi)$, where $V= (d-1)W'^2 - dW^2$ and  $\phi$ approaches an
extremum of $W$ at infinity \cite{Townsend84}.}. We  looked for 
finite mass initial data which evolve to a singularity but do
not have enough mass to enclose the singular region inside a black hole.
We did not find any such examples.
Taken together, these results suggest the big crunch solutions given here
{\em require} boundary conditions that correspond to an unstable dual CFT.
It is then plausible that there will be certain CFT states which do not
have well defined evolution for all time. 
We have seen this happening at the semiclassical level in the deformed $2+1$ 
theory in dual description of our big crunch solutions. 
This is suggestive of the big crunch being 
an endpoint of evolution even in the full string theory.\footnote{It is
possible that the big crunch singularities that we have been
studying in AdS will turn out to have qualitative differences
from true cosmological singularities.}

This would raise the question what determines the boundary conditions at 
cosmological singularities. Perhaps the AdS/CFT correspondence and the 
toy models of cosmologies we have constructed here could be useful to study 
this question further.

\bigskip

\centerline{{\bf Acknowledgments}}
Part of this work was done while G.H. was visiting the IAS in
Princeton and he thanks them for their hospitality. In particular, 
he thanks I. Klebanov, J. Maldacena and N. Seiberg for stimulating
discussions. T.H. thanks the Mitchell Institute for Fundamental Physics 
at Texas A\&M University and the Centro de Estudios Cientificos (CECS) 
in Valdivia, where some of this work was done, for their hospitality.
This work was supported in part by NSF grant PHY-0244764. 

\setcounter{equation}{0}
\appendix

\section{Generalization to arbitrary dimension}

We consider a scalar field with potential $V(\phi)$ coupled to gravity
in $d+1$ dimensions. We assume that $\phi=0$ is a negative local maximum for
the potential (with value corresponding to a unit AdS radius)
and the field has mass $m^2_{BF} < m^2 < m^2_{BF} +1$.
Suppose there is a Euclidean instanton solution of the form
\be \label{instanton}
ds^2 = {d\r^2\over b^2(\r)} +\r^2 d\Omega_d
\ee
and $\phi=\phi(\r)$. Then the field equations determine $b$ in terms of $\phi$
\be \label{bmetric}
b^2(\r) = { 2V \r^2 -d(d-1)\over \r^2 \phi'^2 -d(d-1)}
\ee
The scalar field $\phi$ itself obeys
\be
b^2 \phi'' + \left( {d b^2 \over \r} +bb' \right) \phi' -V_{,\phi} =0
\ee
with $\phi'(0) =0$.
Therefore, asymptotically $\phi$ has the same behavior as the Lorentzian 
scalar field solutions of the same mass  
\be \label{asympt-sc}
\phi = {\alpha\over \r^{\Delta_-}} + {\beta\over \r^{\Delta_+}}
\ee
where 
\be\label{roots}
\Delta_\pm = {d \pm \sqrt{d^2 + 4 m^2}\over 2}
\ee
For instanton solutions, 
the coefficients $\alpha$ and $\beta$ are determined by $\phi(0)$.
Substituting (\ref{asympt-sc}) into (\ref{bmetric}) yields asymptotically
\be\label{asbmetric}
b^2(\r) =  \r^2  +1 +{\Delta_{-} \alpha^2 \over (d-1)}
{1 \over \r^{2\Delta_{-}-2}} +
{4\Delta_{-} \Delta_{+} \alpha \beta \over d(d-1)}{1 \over \r^{d-2}}
\ee

We now compute the euclidean action of the instanton,
\be
I = \int [-{1\over 2} R + {1\over 2}(\nabla \phi)^2 + V(\phi)]
  - \oint K +{1 \over 2d} \oint \left[ (\nabla \phi)^2 -m^2 \phi^2 \right]    
  \ee
where the first surface term is the usual Gibbons-Hawking term, and the
second is a natural generalization of (\ref{charge4d}) to arbitrary dimension.
We will see that for AdS invariant boundary conditions this gives the
surface term required for the Hamiltonian to have well defined functional 
derivatives.
To simplify the calculation of the volume term in the action we can use
a scaling argument similar to the one in section 2. If we rescale
the metric by a constant, the volume term becomes
\be
I(\l^2 g) = \l^{d-1} \int [-{1\over 2} R + {1\over 2}(\nabla \phi)^2]
   + \l^{d+1} \int V(\phi)
   \ee
The instanton must be an extremum of the action, and since the surface 
terms don't contribute to the equations of motion, it must extremize the
volume term. Setting $dI/d\l|_{\l=1} =0$ we learn that
\be
\int [-{1\over 2} R + {1\over 2}(\nabla \phi)^2] =- {d+1\over d-1} \int V(\phi)
\ee
Hence
\be\label{action}
I= -{2\over d-1}\int V(\phi) 
  - \oint K +{1 \over 2d} \oint \left[ (\nabla \phi )^2 -m^2 \phi^2 \right]   
\ee
The divergences can simply be determined by evaluating (\ref{action}) 
using the asymptotic behavior (\ref{asympt-sc})-(\ref{asbmetric}) of 
the instantons. This gives
\beq \label{euclaction}
 -{2\over d-1}\int V(\phi) & = & 
\mathrm{Vol}(S^{d-1})\int^{\r_0} d\r \r^{d-1} 
\left[ d-{d \over 2\r^2} 
-{(2m^2 +d\Delta_-)\alpha^2 \over 2(d-1) \r^{2\Delta_-}} \right]\nonumber\\
- \oint K &= & -d\mathrm{Vol}(S^{d-1})\r^d_0 \left( 1+ {1 \over 2\r^2_0} 
+{\Delta_- \alpha^2 \over 2(d-1) \r_0^{2\Delta_-}}\right)\nonumber\\
{1 \over 2d}\oint \left[ (\nabla \phi )^2 -m^2\phi^2 \right]   & = &
\mathrm{Vol}(S^{d-1}){\Delta_- \alpha^2 \over 2} \r_0^{d-2\Delta_-}
\eeq

The relevant quantity for computing the rate of vacuum decay is the 
difference between the instanton action and the action for pure AdS:
$\Delta I = I - I_{AdS}$. If $\Delta I$ were infinite the probability for 
the vacuum to decay would be zero. 
The background action $I_{AdS}$ diverges as
\be \label{background}
I_{AdS}= d\mathrm{Vol}(S^{d-1}) \left[\int^{\r_0} d\r \r^{d-1} 
\left(1-{1 \over 2\r^2} \right) 
-\r_0^d \left( 1+ {1 \over 2\r_0^2} \right) \right]
\ee
Hence subtracting $I_{AdS}$ removes the leading divergences in $I$, but
since $\phi$ goes to zero so slowly there is a  subleading divergence. 
However, by using (\ref{roots}) one sees the coefficient of the divergent
terms add up to zero. This involves nontrivial cancellations 
among the volume term and both surface terms in the action. A second,
logarithmic divergence coming from $\int V$  does not appear
due to cancellations between the potential and volume element.
The difference $\Delta I = I - I_{AdS}$, therefore, is finite, yielding
a non-zero decay rate.
Note that $\Delta I$ is finite for all $\alpha$ and $\beta$; we have
not yet assumed any relation between them.

If one analytically continues the instanton, one obtains solutions
that evolve to a big crunch.
Since one often uses the boundary conditions (\ref{asympt-sc}) 
to compute correlation functions of operators in the CFT, one might
wonder if all of these theories are unstable. The answer is no. 
If one starts with the standard quantization, $\alpha =0$, then adding
a nonzero $\alpha$ corresponds to adding a term  proportional
to $\alpha$ to the CFT action. 
In this context one usually assumes $\alpha$ is a constant, but we
have seen in section 3.5 that the analytic continuation of the
instanton corresponds to
an $\alpha$ which is time dependent and diverges in finite time. So using
the standard boundary conditions, 
the big crunch would correspond to adding an
explicitly diverging term to the action.
In addition, once one modifies the CFT
action, the usual vacuum $AdS_{d+1}$
is no longer a solution since the field is required to
satisfy (\ref{asympt-sc}) with nonzero $\alpha$.

By contrast, AdS invariant boundary conditions exist in all dimensions 
\cite{Hertog04}. Asymptotic AdS invariance only requires
$\beta = f \alpha^{\Delta_{+}/\Delta_{-}}$, where $f$ is an arbitrary constant.
So $\alpha$ and $\beta$ are related, but can be zero or time 
dependent. In general there is a contribution from the scalar 
field to the conserved charges,
\be
\label{gencharge}
Q[\xi]=Q_G[\xi]+{1 \over 2d} \oint \xi^\perp 
\left[ (\nabla \phi )^2 -m^2 \phi^2 \right]    
\ee
where $Q_G[\xi]$ is the standard gravitational surface term (\ref{gravcharge}).
For spherical solutions this gives
\be \label{genmass}
Q[\partial_t]=  \mathrm{Vol}(S^{d-1})\left( {d-1 \over 2} M_0 
- {2f m^2 \alpha^{d/\Delta_{-}} \over d} \right)
\ee
where $M_0$ is the coefficient of the $O(1/r^{d+2})$ correction to the
$g_{rr}$ component of the AdS metric. Hence the gravitational contribution 
$Q_G[\partial_t]$ to the mass of the instantons can be read 
off from (\ref{asbmetric}). One sees it diverges as $\rho^{d-2\Delta_{-}}$.
However this divergence, as well as the finite gravitational contribution 
proportional to $M_0$, are exactly cancelled by the scalar contributions. 
Instantons of this type thus specify nontrivial, zero mass initial data for 
all $m^2$ in the range $m^2_{BF} < m^2 < m^2_{BF} +1$ and in all dimensions 
$d+1$. 
In general, the CFT duals of these  AdS invariant boundary conditions
are not yet well understood. We discussed the simplest case in section 3.

Finally we turn to the case of a scalar field with $m^2 = m^2_{BF}$.
For fields that saturate the BF bound, $\Delta_+ = \Delta_- =d/2$ 
and the second solution asymptotically behaves like $\ln \r/\r^{d/2}$.
Thus asymptotically we have
\be \label{asympt-sc2}
\phi = {\alpha \ln \r\over \r^{d/2}} + {\beta\over \r^{d/2}}
\ee
and therefore
\be \label{asb5d}
b^2=  \r^2  +1 +{ d\alpha^2 (\ln \r)^2 \over 2(d-1) \r^{d-2}}
+{\alpha(d \beta -\alpha) \ln \r \over (d-1) \r^{d-2}}
+{d^2 \beta^2 -2d \alpha \beta +2\alpha^2 \over 2d(d-1) \r^{d-2}}
\ee
Asymptotic solutions of this form can similarly be regarded as having
AdS invariant boundary conditions provided $\beta = \alpha(f-
{2\over d} \ln \alpha)$ \cite{Hertog04,Henneaux04}.
As before, the scalar surface term (\ref{gencharge}) in the Hamiltonian gives
in general a non-vanishing contribution to the conserved charges.
For instance, the total mass of spherically symmetric solutions is given by 
\beq \label{genmass5d}
Q[\partial_{t}] &=& \mathrm{Vol}(S^{d-1})\left[{d-1 \over 2}
M_0+\frac{1}{d}\alpha^2(\ln \alpha)^2
+\alpha^2\left(\frac{1}{d}-f\right)\ln \alpha  \right.\nonumber\\
& & \left. +
{ \alpha^2 \over 2}\left({d \over 2} f^2-f+\frac{1}{d}\right)
\right].  
\eeq
It is easily seen from (\ref{asb5d}) that the instantons have zero mass.
Evaluating the action (\ref{action}) gives
\beq
 -{2\over d-1}\int V(\phi)  & = & \mathrm{Vol}(S^{d-1})
\int^{\r_0} { d\r \over \r} \left[d\r^d-{d \over 2}\r^{d-2} 
+{d\alpha^2 \ln  \r + d\alpha \beta -\alpha^2 
\over 2(d-1)} \right]
\nonumber\\
- \oint K  & = & -\mathrm{Vol}(S^{d-1}) \left( d \r^d_0 + 
{d \over 2}\r^{d-2}_0 
+{ d^2\alpha^2 \ln^2 \r_0 +2d \alpha(d\beta- \alpha) \ln \r_0 
\over 4(d-1)} \right) \nonumber\\
\oint \left[ 
{d \over 8}\phi^2 + {(\nabla \phi)^2 \over 2d} \right]   
& = & { \mathrm{Vol}(S^{d-1}) \over 4} \left[ d\alpha^2  \ln^2 \r_0 +
2(d\alpha \beta- \alpha^2 )\ln \r_0 \right]
\eeq
Subtracting the background action (\ref{background}) and using 
(\ref{roots}) again yields a finite result. The value of the Euclidean action 
determines the rate of vacuum decay. The analytic continuation of the
instantons again provides examples of smooth initial data
which evolve to a big crunch. 


\begin{thebibliography}{99}

\bibitem{Hartle83}
J. B. Hartle, S. W. Hawking,
``The Wave Function of the Universe,''
Phys. Rev. {\bf D28} (1983) 2960

\bibitem{Gasperini93}
M. Gasperini, G. Veneziano, ``Pre-Big Bang in String Cosmology,''
Astropart. Phys. {\bf 1} (1993) 317, hep-th/9211021;
``The pre-big bang scenario in string cosmology,''
Phys.\ Rept.\  {\bf 373} (2003) 1, hep-th/0207130

\bibitem{Khoury02}
J. Khoury, B. A. Ovrut, N. Seiberg, P.J. Steinhardt, N. Turok,
``From Big Crunch to Big Bang,''
Phys. Rev. {\bf D65} (2002) 086007, hep-th/0108187;
 P.J. Steinhardt, N. Turok, ``Cosmic Evolution in a Cyclic Universe,''
Phys. Rev. {\bf D65} (2002) 126003, hep-th/0111098

 
\bibitem{Maldacena98}
J. M. Maldacena,
``The large N limit of superconformal field theories and supergravity,''
Adv.\ Theor.\ Math.\ Phys.\  {\bf 2} (1998) 231, hep-th/9711200

\bibitem{Fidkowski04}
L. Fidkowski, V. Hubeny, M. Kleban, S. Shenker,
``The Black Hole Singularity in AdS/CFT,''
JHEP {\bf 0402} (2004) 014, hep-th/0306170

\bibitem{Hertog04}
T. Hertog, K. Maeda,
``Black Holes with Scalar Hair and Asymptotics in $N=8$ Supergravity,''
hep-th/0404261

\bibitem{Breitenlohner82}
P.~Breitenlohner and D.~Z.~Freedman, ``Stability In Gauged Extended 
Supergravity,'' Annals Phys.\  {\bf 144} (1982) 249;
``Positive Energy In Anti-De Sitter Backgrounds And Gauged Extended
Supergravity,'' Phys.\ Lett.\ B {\bf 115} (1982) 197

\bibitem{Henneaux04}
M. Henneaux, C. Martinez, R. Troncoso, J. Zanelli,
``Asymptotically Anti-de Sitter Spacetimes and Scalar Fields with a 
Logarithmic Branch,'' hep-th/0404236

\bibitem{Hertog03b}
T. Hertog, G. T. Horowitz, K. Maeda, 
``Negative Energy in String Theory and Cosmic Censorship Violation,''
Phys.\ Rev.\ D {\bf 69}, 105001 (2004), hep-th/0310054

\bibitem{Gutperle04}
M. Gutperle and P. Kraus,
``Numerical Study of Cosmic Censorship in String Theory,''
JHEP {\bf 0404} (2004) 024, hep-th/0402109

\bibitem{Hubeny:2004cn}
V.~E.~Hubeny, X.~Liu, M.~Rangamani and S.~Shenker,
``Comments on cosmic censorship in AdS/CFT,'' hep-th/0403198

\bibitem{Maldacena:1998uz}
J.~M.~Maldacena, J.~Michelson and A.~Strominger,
``Anti-de Sitter fragmentation,''
JHEP {\bf 9902} (1999) 011, hep-th/9812073

\bibitem{Heusler92}
M. Heusler, N. Straumann,
``Scaling Arguments for the Existence of Static, Spherically Symmetric 
Solutions of Self-Gravitating Systems,''
Class. Quant. Grav. {\bf 9} (1992) 2177;
M.~Heusler,
``A No Hair Theorem For Selfgravitating Nonlinear Sigma Models,''
J.\ Math.\ Phys.\  {\bf 33} (1992) 3497

\bibitem{Sudarsky:ty}
D.~Sudarsky and R.~M.~Wald,
``Extrema Of Mass, Stationarity, And Staticity, And Solutions To The Einstein
Yang-Mills Equations,''
Phys.\ Rev.\ D {\bf 46} (1992) 1453

\bibitem{deWit82}
B. de Wit, H. Nicolai,
``$N=8$ Supergravity with Local $SO(8) \times SU(8)$ Invariance,''
Phys. Lett. {\bf 108B} (1982) 285; ``$N=8$ Supergravity,''
Nucl. Phys. {\bf B208} (1982) 323

\bibitem{Duff99}
M. J. Duff, J. T. Liu,
``Anti-de Sitter Black Holes in Gauged N=8 Supergravity,''
Nucl. Phys. {\bf B554} (1999) 237, hep-th/9901149

\bibitem{Abbott82}
L. F. Abbott, S. Deser,
``Stability of Gravity with a Cosmological Constant,''
Nucl. Phys. {\bf B195} (1982) 76

\bibitem{Henneaux85}
M. Henneaux, C. Teitelboim,
``Asymptotically Anti-de Sitter Spaces,''
Comm. Math. Phys. {\bf 98} (1985) 391

\bibitem{Klebanov99}
I.R. Klebanov, E. Witten, 
``AdS/CFT Correspondence and Symmetry Breaking,''
Nucl. Phys. {\bf B556} (1999) 89, hep-th/9905104

\bibitem{deHaro:2000xn}
S.~de Haro, S.~N.~Solodukhin and K.~Skenderis,
``Holographic reconstruction of spacetime and renormalization in the  AdS/CFT
correspondence,''
Commun.\ Math.\ Phys.\  {\bf 217} (2001) 595, hep-th/0002230

\bibitem{Bianchi02}
M. Bianchi, D. Z. Freedman, K. Skenderis,
``Holographic Renormalization,''
Nucl. Phys. {\bf B631} (2002) 159, hep-th/0112119

\bibitem{Dafermos:2004ws}
M.~Dafermos,
``A note on naked singularities and the collapse of self-gravitating Higgs
fields,'' gr-qc/0403033

\bibitem{Coleman:1980aw}
S.~R.~Coleman and F.~De Luccia,
``Gravitational Effects On And Of Vacuum Decay,''
Phys.\ Rev.\ D {\bf 21} (1980) 3305

\bibitem{Aharony:1998rm}
O.~Aharony, Y.~Oz and Z.~Yin,
``M-theory on AdS(p) x S(11-p) and superconformal field theories,''
Phys.\ Lett.\ B {\bf 430} (1998) 87, hep-th/9803051

\bibitem{Witten02}
E. Witten, ``Multi-Trace Operators, Boundary Conditions, and AdS/CFT 
Correspondence,'' hep-th/0112258

\bibitem{Gubser:2002vv}
S.~S.~Gubser and I.~R.~Klebanov,
``A universal result on central charges in the presence of double-trace
deformations,''
Nucl.\ Phys.\ B {\bf 656} (2003) 23, hep-th/0212138

\bibitem{Gubser03}
S. Gubser, I. Mitra,
``Double-Trace Operator and One-Loop Vacuum Energy in AdS/CFT,''
Phys. Rev. {\bf D67} (2003) 064018, hep-th/0210093

\bibitem{Gunaydin85}M.~Gunaydin, L.~J.~Romans, N.~P.~Warner, 
``Gauged $N=8$ Supergravity in Five Dimensions,'' Phys. Lett. 
{\bf 154B} (1985) 268;
``Compact and Noncompact Gauged Supergravity Theories in Five Dimensions,''
Nucl. Phys. {\bf 272} (1986) 598

\bibitem{Pernici85}M.~Pernici, K.~Pilch, P. van Nieuwenhuizen,
``Gauged $N=8$ $D=5$ Supergravity,''Nucl. Phys. {\bf B259} (1985) 460

\bibitem{Freedman:1999gk}
D.~Z.~Freedman, S.~S.~Gubser, K.~Pilch and N.~P.~Warner,
``Continuous distributions of D3-branes and gauged supergravity,''
JHEP {\bf 0007} (2000) 038, hep-th/9906194

\bibitem{Gibbons83}
G.~W.~Gibbons, C.~M.~Hull and N.~P.~Warner,
``The Stability Of Gauged Supergravity,''
Nucl.\ Phys.\ B {\bf 218} (1983) 173

\bibitem{Hertog03}
T.~Hertog, G.~T.~Horowitz, K.~Maeda,
``Generic Cosmic Censorship Violation in anti de Sitter Space,''
Phys. Rev. Lett. {\bf 92} (2004) 131101, gr-qc/0307102

\bibitem{Hertog04b}
T.~Hertog, G.~T.~Horowitz, K.~Maeda,
``Update on Cosmic Censorship in AdS,'' gr-qc/0405050

\bibitem{Townsend84}
P.~K.~Townsend,
``Positive Energy And The Scalar Potential In Higher Dimensional 
(Super)Gravity Theories,''
Phys.\ Lett.\ B {\bf 148} (1984) 55



\end{thebibliography}
\end{document}